\documentclass[conference]{IEEEtran}

\IEEEoverridecommandlockouts
\usepackage{cite}
\usepackage{amsmath,amssymb,amsfonts}
\usepackage{algorithmic}
\usepackage{algorithm}
\usepackage{graphicx}
\usepackage{subfig}
\usepackage{textcomp}
\usepackage{xcolor}
\usepackage{booktabs}
\usepackage[font=small, belowskip=-2pt]{caption}
\usepackage{url}

\def\BibTeX{{\rm B\kern-.05em{\sc i\kern-.025em b}\kern-.08em
    T\kern-.1667em\lower.7ex\hbox{E}\kern-.125emX}}

\newcommand{\CO}{$\mathrm{CO_2}$}
\begin{document}

\title{SciAI4Industry - Solving PDEs for industry-scale problems with deep learning 
}

\author{\IEEEauthorblockN{1\textsuperscript{st} Philipp A. Witte}
\IEEEauthorblockA{\textit{Microsoft}\\
Redmond, USA \\
pwitte@microsoft.com} \\
\and
\IEEEauthorblockN{2\textsuperscript{nd} Russell J. Hewett}
\IEEEauthorblockA{\textit{Microsoft}\\
Redmond, USA \\
rhewett@microsoft.com} \\
\and
\IEEEauthorblockN{3\textsuperscript{rd} Kumar Saurabh}
\IEEEauthorblockA{\textit{Iowa State University}\\
Ames, USA \\
maksbh@iastate.edu} \\
\and
\IEEEauthorblockN{4\textsuperscript{th} AmirHossein Sojoodi}
\IEEEauthorblockA{\textit{Queens University} \\
Kingston, Canada \\
amir.sojoodi@queensu.ca} \\
\and
\IEEEauthorblockN{5\textsuperscript{th} Ranveer Chandra}
\IEEEauthorblockA{\textit{Microsoft}\\
Redmond, USA \\
ranveer@microsoft.com} \\
}

\maketitle

\begin{abstract}

Solving partial differential equations with deep learning makes it possible to reduce simulation times by multiple orders of magnitude and unlock scientific methods that typically rely on large numbers of sequential simulations, such as optimization and uncertainty quantification. Two of the largest challenges of adopting scientific AI for industrial problem settings is that training datasets must be simulated in advance and that neural networks for solving large-scale PDEs exceed the memory capabilities of current GPUs. We introduce a distributed programming API in the Julia language for simulating training data in parallel on the cloud and without requiring users to manage the underlying HPC infrastructure. In addition, we show that model-parallel deep learning based on domain decomposition allows us to scale neural networks for solving PDEs to commercial-scale problem settings and achieve above 90\% parallel efficiency. Combining our cloud API for training data generation and model-parallel deep learning, we train large-scale neural networks for solving the 3D Navier-Stokes equation and simulating 3D \CO{} flow in porous media. For the \CO{} example, we simulate a training dataset based on a commercial carbon capture and storage (CCS) project and train a neural network for \CO{} flow simulation on a 3D grid with over 2 million cells that is 5 orders of magnitudes faster than a conventional numerical simulator and 3,200 times cheaper.

\end{abstract}


\section{Motivation \& Objectives}
\label{section:motivation}

Solving partial differential equations (PDEs) with numerical methods plays an important role in many industrial fields such as aerodynamic shape design, exploration seismology, finance, carbon capture and storage (CCS), or renewable energies. Commercial and open-source simulation packages are conventionally based on the finite difference (FD), finite volume (FV), or finite element method (FEM), but recently there has been a growing interest in solving PDEs with various machine or deep learning (ML/DL) methods \cite{karniadakis2021,lavin2021}. Deep learning in the context of numerical simulations, which falls under the umbrella of \textit{scientific AI/ML} or \textit{SciML}, promises to reduce the simulation time of PDEs by several orders of magnitude compared to traditional solvers \cite{hennigh2021nvidia} or simulate phenomena for which the underlying PDE is unknown \cite{udrescu2020ai}. These factors make deep learning-based approaches attractive for applications that require many sequential simulations such as inverse problems and uncertainty quantification (UQ) \cite{lavin2021}.

For many commercial-scale applications, numerical simulators must be able to solve time-dependent PDEs on large-scale meshes with millions of grid points and therefore most simulation packages use techniques from high-performance computing (HPC) to scale to large-scale problem sizes on HPC clusters. Current state-of-the-art methods for solving PDEs with deep learning however, have so far been limited to either 2D problem sizes or small-scale 3D problems, with typical mesh sizes that lie below or around one million grid points \cite{botelho2020deep,  pathak2022fourcastnet}. The main reason for this limitation is the amount of available GPU memory, as memory demand for training neural networks scales with the size of the input and output data. Solving PDEs with DL at large problem sizes beyond the memory capacity of a single GPU requires model- rather than data parallelism, the latter being currently the most widely used form of parallelism in deep learning \cite{ben2019demystifying}.

A second important challenge of training large-scale deep surrogate models is the simulation of training data. In scenarios where we are interested in training networks for solving PDEs that generalize to different boundary/initial conditions or sets of PDE coefficients (e.g., material or control parameters), scientific AI approaches are based on supervised learning \cite{lavin2021}. Training data for supervised learning consists of pairs of input data (boundary/initial conditions, PDE coefficients) and output data (solutions of the PDE as a function of space and time) and therefore requires running many conventional numerical simulations prior to training. For large-scale industrial applications as reservoir simulation, users must run 100s to 1,000s of simulations for training data generation, each of which potentially take multiple hours to run on a multi-core machine \cite{wen2021u}. Simulating training data for scientific ML applications therefore requires access to HPC infrastructure, which is only available to a limited number of researchers (typically at national and corporate labs). Cloud computing offers an alternative to on-premise HPC clusters and is publicly available, thereby providing an opportunity to democratize access to HPC infrastructure. However, running HPC workloads in the cloud involves significant administrative challenges, as users are responsible for creating and managing virtual HPC clusters themselves. This leads to a significant amount of complexity that makes it difficult to scale deep-learning based surrogate models for solving PDEs to industry-scale problem sizes.

This work aims to address these two outlined challenges and scale deep neural networks for solving PDEs to industry-scale problem sizes. To achieve this, our main objectives are:

\begin{enumerate}

\item Develop a software package that simplifies running HPC workloads in the cloud, with an emphasis on simulating training data for scientific AI workloads.

\item Discuss why current approaches to parallelism in deep learning are inadequate for scaling scientific AI models to industrial-scale applications and show that model parallelism based domain decomposition is a more promising approach that achieves high levels of parallel efficiency (above 90\%).

\item Demonstrate that addressing the above challenges enables us to apply scientific AI to a real-world reservoir simulation problem and train the largest deep learning-based numerical simulator to date.

\end{enumerate}

\section{Related work}

\subsection{Scientific AI for industry applications}

Our work is motivated by recent advances in scientific AI on solving PDEs with deep learning and in particular on its application to industrial problems such as shape optimization \cite{li2021deep, hennigh2021nvidia}, weather and climate forecasting \cite{salman2015weather, kurth2022fourcastnet} or computational chemistry \cite{butler2018machine, goh2017deep}. We are particularly interested in numerical reservoir simulations for simulating subsurface \CO{} flow in the context of carbon capture and storage (CCS). Simulating \CO{} flow in porous media involves solving coupled systems of non-linear equations with implicit numerical solvers and thus holds a large potential for improving simulation speeds with deep learning. Several deep learning-based \CO{} flow simulators have been introduced in recent years, most of which are 2D simulators \cite{mo2019deep, shokouhi2021physics, zhong2019predicting, jiang2021deep, tang2021adeep, wen2021u}, but including several 3D simulators as well \cite{tang2021deep, yan2021robust}. Even though some of these DL-based simulators are advertised as \textit{large scale} \cite{jiang2021deep} or \textit{commercial scale} \cite{tang2021adeep}, the problem sizes considered are considerably smaller than typical problem sizes encountered in production settings. The largest network from \cite{tang2021deep} predicts \CO{} flow on a 3D mesh with 133,000 cells over 10 time steps, which is an order of magnitude smaller than open-source \CO{} benchmarks such as the Sleipner model (2.1 million grid cells). For other applications, the largest AI simulator trained on GPUs, to the best of our knowledge, is the U-Net from \cite{balu2021distributed} for solving the 3D Poisson equation on a $256^3$ grid (16 million predicted variables).

\subsection{Parallelism in deep learning}

\begin{figure}[bt]
\centerline{\includegraphics[width=0.99\columnwidth]{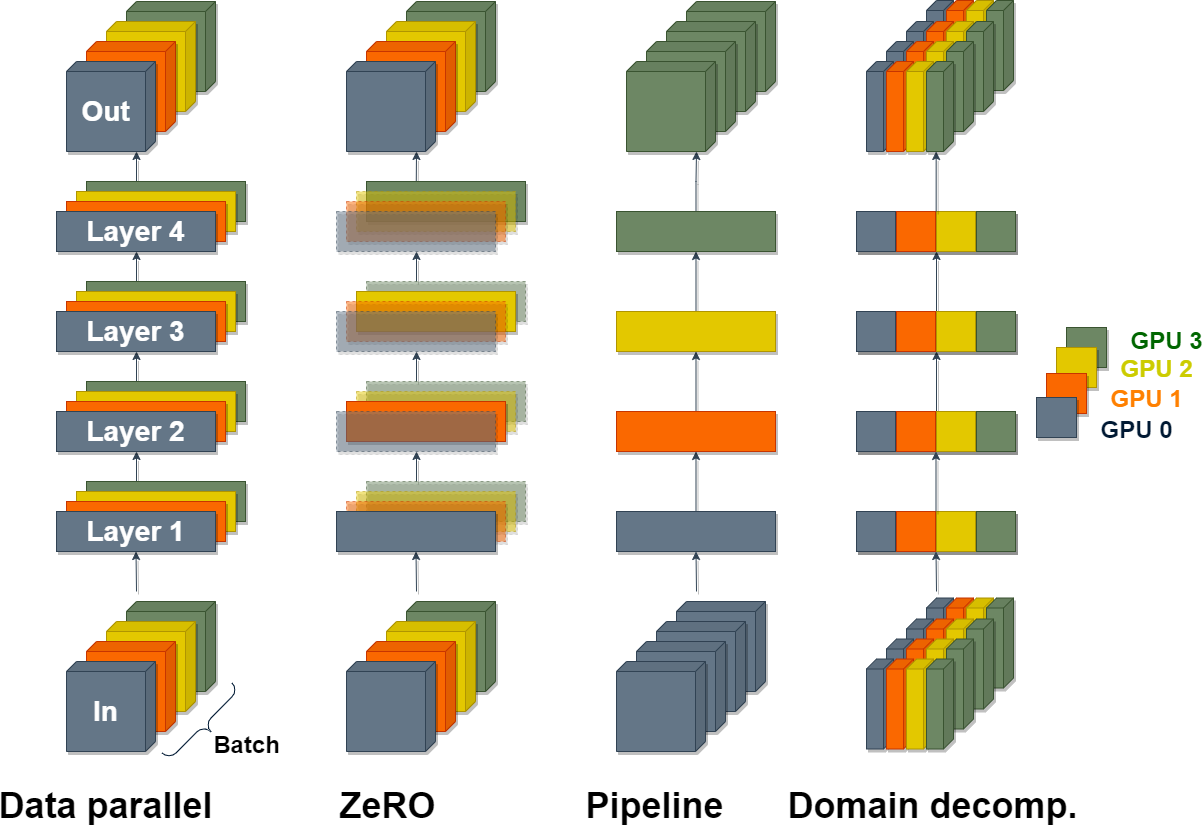}}
\caption{Different strategies for parallelizing deep neural networks with data or model parallelism.}
\label{fig:parallel-strategies}
\end{figure}

Current AI-based simulators are not able to scale to larger problem sizes, because they are trained with data parallelism, which is the most widely available form of parallelism and supported by all major DL frameworks \cite{paszke2019pytorch, abadi2016tensorflow, frostig2018compiling}. In data parallelism (Fig.~\ref{fig:parallel-strategies}), samples from a batch of data are partitioned across multiple GPUs, but each GPU must be able to fit at least one data sample (including its hidden states), as well as network weights and gradients into memory \cite{ben2019demystifying}. The Zero Redundancy Optimizer (ZeRO) from DeepSpeed \cite{rasley2020deepspeed} has enabled the training of large natural language processing (NLP) models such as GPT3 \cite{brown2020language} by removing redundant copies of the network across GPUs that data parallelism induces. ZeRO partitions network weights across GPUs but still requires that each GPU stores the hidden states (activations) of at least one data sample. Distributing network weights rather than the hidden states of the data is advantageous for NLP models based on the transformer architecture, whose memory footprint is dominated by those weights, but is less effective for architectures such as convolutional neural networks (CNNs) that are common in scientific AI.

Aside from data parallelism, the other most widely supported form of parallelism in deep learning is pipeline parallelism, in which layers of the networks are partitioned across multiple GPUs \cite{ben2019demystifying} and data is processed sequentially by each GPU. While pipeline parallelism distributes both network weights and hidden states, it relies on large batch sizes to achieve high concurrency \cite{narayanan2019pipedream, huang2019gpipe}. \textit{Domain decomposition} offers an approach to parallelism for neural networks that does not rely on any particular batch size or network architecture for concurrency. In domain decomposition, both the input data (including hidden states), and network parameters are distributed, which is why it is often referred to as tensor parallelism. So far, tensor parallelism has been mainly applied to transformer networks in the context of NLP with networks such as Megatron \cite{shoeybi2019megatron} or the Turning Natural Language Generation (NLG) model \cite{smith2022using}. The authors in \cite{grady2022towards} are the first to apply tensor parallelism to scientific AI by implementing a model-parallel version of the Fourier Neural Operators architecture \cite{li2020fourier}. The implementation in \cite{grady2022towards} is based on DistDL, a package that provides tensor decomposition and parallel communication primitives for Pytorch \cite{hewett2020linear}. In scientific AI, domain decomposition is also used to describe multiple individual networks that are each trained to predict a subset of the total output \cite{li2019d3m, jagtap2021extended}. This form of domain decomposition however requires constrained optimization to enforce continuity along domain boundaries and may not lead to consistent results.

\subsection{HPC in the cloud}

Existing frameworks for running HPC workloads in the cloud can be grouped into \textit{traditional} cluster managers and cloud-native cluster managers. The former category includes services such as Azure Cycle Cloud and AWS ParallelCluster, which enable users to spin up HPC clusters that resemble traditional on-premise systems with distributed network file systems and HPC schedulers such as SLURM and PBS. Cloud-native approaches include services such as Kubernetes \cite{kubernetes2021}, AWS/Azure/GCP Batch \cite{awsbatch2021, azurebatch2021}, which are typically based on containerization, object storage and first party job schedulers managed by cloud platforms. Both approaches involve high levels of complexity in selecting the appropriate hardware configuration and target primarily HPC administrators rather than end users. Similarly, even frameworks that are more geared towards data scientists such as Hadoop \cite{shvachko2010hadoop} and Spark \cite{zaharia2010spark}, require upfront configurations of clusters that users can connect to.

Serverless function frameworks such as Azure Functions, GPC Functions, or AWS Lambda offer the possibility to run code in the cloud without requiring the user to manage the underlying compute infrastructure \cite{awslambda2022, azurefunctions2022, gcpfunctions2022}. However, serverless computing is not geared towards HPC, as it does not allow users to specify hardware (e.g., CPU architectures or GPUs) and has restrictions on maximum allowed run-time (e.g., 15 minutes on AWS) and available memory (e.g., 10 GB). Several projects have shown that it is possible to run certain HPC workloads on top of serverless functions frameworks by decomposing workloads into small portions \cite{fouladi2017encoding, shankar2018numpywren, zhang2020kappa}, but they are custom solutions that do not translate to arbitrary applications and do not enable users to run third-party simulators on top of serverless functions.

Our goal is to enable users to execute long-running simulators on the cloud that are either manually implemented or based on third party simulators such as the Open Porous Media simulator \cite{rasmussen2021open}, without having to manually manage the underlying HPC infrastructure. We achieve this goal through a distributed programming package in the Julia programming language that is built on top of Azure Batch. The user API resembles other task-based distributed programming packages such as Dask \cite{rocklin2015dask} and Ray \cite{moritz2018ray}, but through its tight integration with Azure Batch, does not require users to mange the underlying infrastructure.

\section{Key insights \& contributions}

To scale scientific AI to industrial-scale problem sizes, we must overcome two challenges. First, we must enable users without access to traditional HPC clusters to generate simulated training data.  Second, we must enable scientific AI training for neural network models and high-dimensional scientific data sets with millions or billions of degrees-of-freedom. To achieve the former, we demonstrate that batch computing services such as Azure Batch satisfy the necessary requirements for running large-scale HPC workloads in the cloud and that they can be made accessible to scientists through abstractions that expose these services as distributed programming frameworks to the user. To solve the second challenge, our main insight is that domain decomposition achieves much better levels of concurrency and scaling than alternate model parallel approaches, and even a relatively small number of GPUs suffices for training industrial-scale simulators. Our two main contributions that enable the scalability of scientific AI to industry-scale problems are summarized as follows:

\begin{itemize}

\item We introduce \textit{Redwood.jl}, an open-source Julia package for running scientific computing workloads in the cloud without having to manage the underlying infrastructure. Our package enables users to run both Julia and third party simulators on the cloud for simulating data in the context of scientific AI.

\item We show that pipeline parallelism is not well suited for training an FNO-based AI simulator, but we can reach above 90 percent parallel efficiency (on up to 8 Nvidia A100 GPUs) with domain decomposition. We achieve this by improving the parallel FNO implementation from \cite{grady2022towards} by adding support for NCCL \cite{jeaugey2017nccl} and reducing overall communication volume. 

\item Leveraging these contributions, we train the largest surrogate models for solving PDEs to date. In the first example, we train an FNO for simulating turbulent flow around a sphere on a spatial-temporal grid of 130 $\times$ 130 $\times$ 130 $\times$ 84 grid points (140 million solution points, in total) using 3,200 simulated training samples. In our second example, we train an FNO for simulating \CO{} flow on the Sleipner geomodel, a real-world reservoir simulation benchmark from the world's first industrial CCS project. We simulate 1,600 training examples and train an FNO to predict \CO{} flow on the original simulation grid of 262 $\times$  118 $\times$ 64 grid points for 86 time steps - a total of 170 million predicted variables and an order of magnitude larger than the current largest AI simulator trained on GPUs from \cite{balu2021distributed}. 

\end{itemize}

\section{Architecture and performance evaluation}

\begin{figure}[bt]
\centerline{\includegraphics[width=0.99\columnwidth]{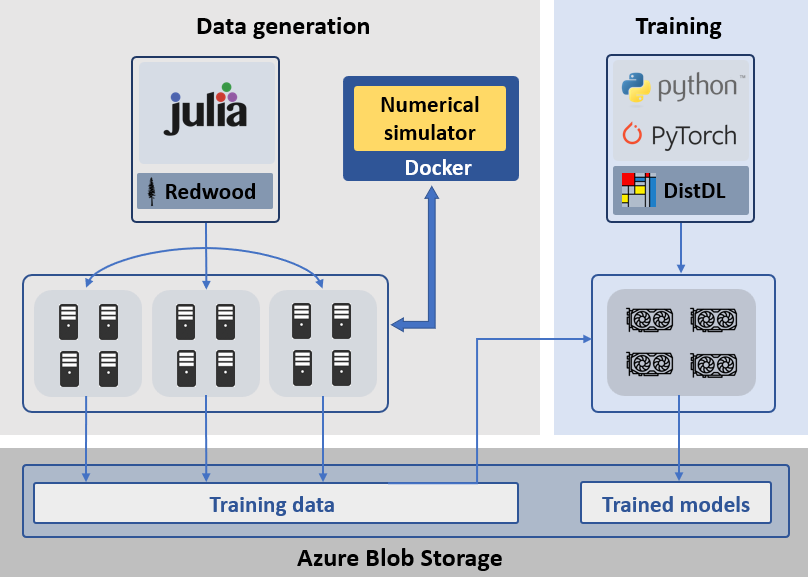}}
\caption{Architecture for training industry-scale AI-driven numerical simulators. The data generation component is based on Redwood, a Julia API for running existing numerical simulators on up to thousands of virtual machines on Azure.}
\label{fig:architecture}
\end{figure}

Our architecture for scaling scientific AI to industry-scale applications has two components: an API for parallel training data generation in the cloud and a model-parallel FNO implementation based on DistDL (Fig.~\ref{fig:architecture}). Our contribution for the data generation component is Redwood, a distributed programming framework in the Julia language that enables users to run simulators written in Julia or binary code on the Azure cloud. We choose the Julia programming language for this component over Python, because Julia is designed from the ground up for numerical computing with an emphasis on high performance and multi-platform support via just-in-time compilation \cite{bezanson2017julia}. Our model-parallel FNO implementation is written in Python and based on the implementation described in \cite{grady2022towards}. We further optimize the performance for scalability on a single Nvidia DGX (up to eight A100s) by adding NCCL support to DistDL and by reducing data communication in the FNO implementation. The next two sections describe each architecture component in more detail, starting with the Julia framework for parallel training data generation on Azure.

\subsection{Redwood architecture}

Redwood is a distributed programming framework built on top of Azure's first-party batch computing service \textit{Azure Batch}. The idea of Redwood is to relieve users from managing HPC infrastructure on the cloud, while at the same time preventing users from having to interact with platform- or cloud-specific user/REST APIs. Instead, users interact with Redwood's distributed programming macros that closely resemble Julia's existing macros for cluster-based HPC. By cluster based, we mean that conventionally, users first need to administer an HPC cluster with cloud services such as Azure Cycle Cloud or Kubernetes and then use an HPC scheduler such as SLURM or PBS to request parallel resources on the cluster. 

\begin{figure}[bt]
\subfloat{\label{fig:redwood-code-a}}{\includegraphics[width=0.49\columnwidth]{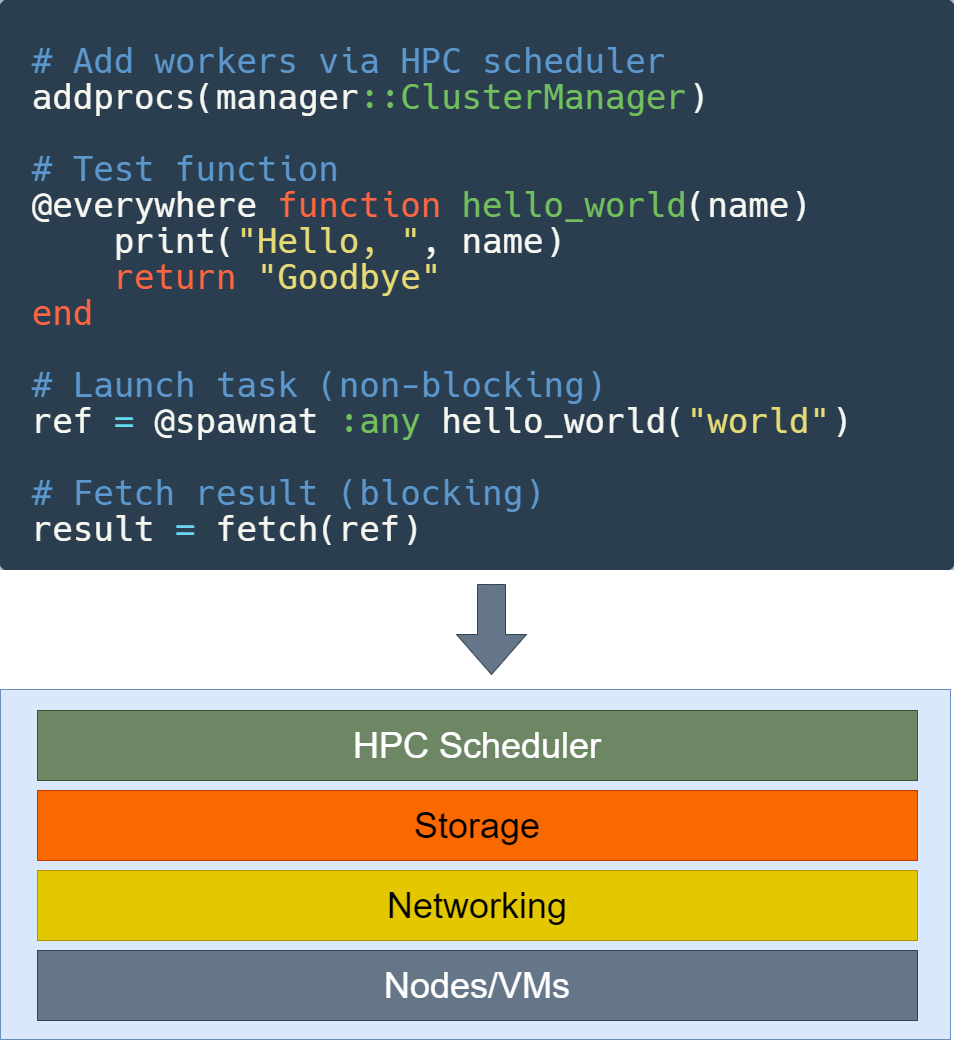}}
\subfloat{\label{fig:redwood-code-b}}{\includegraphics[width=0.49\columnwidth]{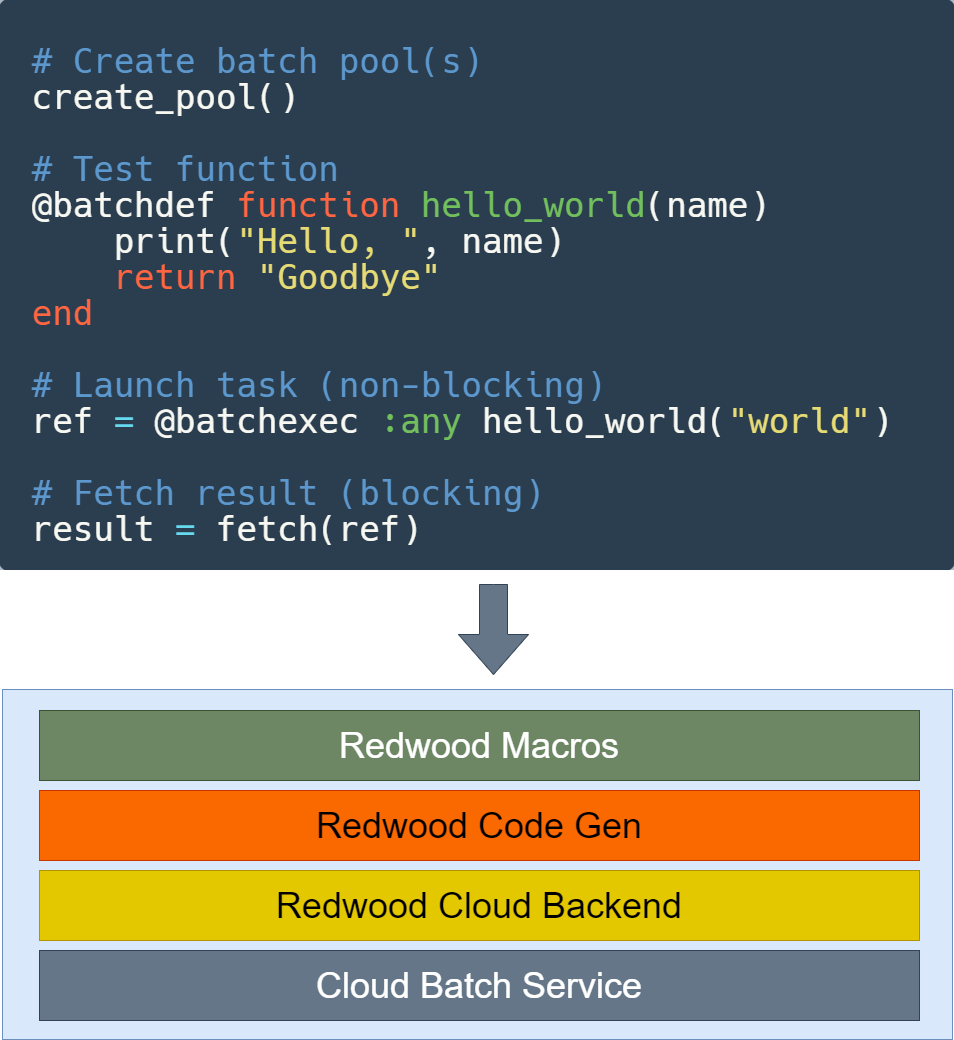}}
\caption{Running a hello-world example with Julia's conventional distributed programming model on an HPC cluster and clusterless HPC with Redwood.}
\label{fig:redwood-code}
\end{figure}

Julia's native distributed programming framework is primarily based on task parallelism using one-sided communication statements. The main primitives that enable this style of communication are remote functions calls and remote references. To remotely execute a function on parallel workers, users first tag their function with the \texttt{@everywhere} macro, which makes the function known to the parallel workers and then execute it via the \texttt{@spawnat} macro. This macro executes the code on a specified remote worker and returns a reference to its function output, which can be copied to the master by calling the \texttt{fetch} function (Fig.~\ref{fig:redwood-code-a}).

Redwood provides analogous macros and functions for executing Julia code through Azure Batch. This means that instead of running a parallel Julia session on top of a (user-managed) HPC cluster, users execute remote functions calls from their laptop or a single cloud node through Azure Batch. The main difference to the conventional approach is that the main Julia program is not connected to any of the worker nodes directly and instead, remote function calls are scheduled and executed via Azure Batch. Redwood provides macros for remotely executing functions on one or multiple workers, for fetching remote references, as well as for broadcasting variables. This makes it possible to convert a conventional distributed Julia program to one that runs on top of Azure Batch with minor changes to the code (Fig.~\ref{fig:redwood-code-b}). The current Redwood version supports Azure Batch only, but in principle, adding additional backends (e.g., for AWS or GCP) is possible as well.

Redwood's core functionality is the execution of tagged Julia functions as parallel Azure Batch jobs and/or tasks. The \texttt{@batchexec} macro creates a closure around the executed expression, serializes the function's abstract syntax tree (AST), and submits a batch job to Azure using the Azure Batch user API (which the Redwood user never interacts with directly). More specifically, calling a function with the \texttt{@batchexec} macro involves the following steps: (1) parsing of function input arguments, (2) splitting of expressions into parallel tasks (for more than one task), (3) replacing of \texttt{return} statements with the serialization of output arguments to object storage, (4) serializing the ASTs of previously tagged expressions and of the executed expression and uploading them to cloud storage, (5) making API calls to create batch jobs/tasks, (6) returning a control structure with a reference to the (future) function output.

The remote Azure Batch workers each run a light-weight Redwood runtime, which downloads and de-serializes the uploaded ASTs and compiles and runs them on the local architecture. By default, Redwood executes functions on the smallest Azure virtual machines (VMs), but users can specify any other VM types that are supported by Azure Batch, including the HBv3 series with InfiniBand interconnect, 120 CPU cores and 448 GB of memory that specifically targets HPC workloads. Redwood's default behavior is to execute remote function calls as individual tasks that each run on a single node and cannot communicate with each other. However, users can also enable multi-node parallelism and execute function calls that run across multiple VMs, e.g., by combining Redwood with Julia's MPI interface.

\subsection{Redwood performance}

\begin{figure}[bt]
\subfloat{\label{fig:redwood-scaling-a}}{\includegraphics[width=0.49\columnwidth]{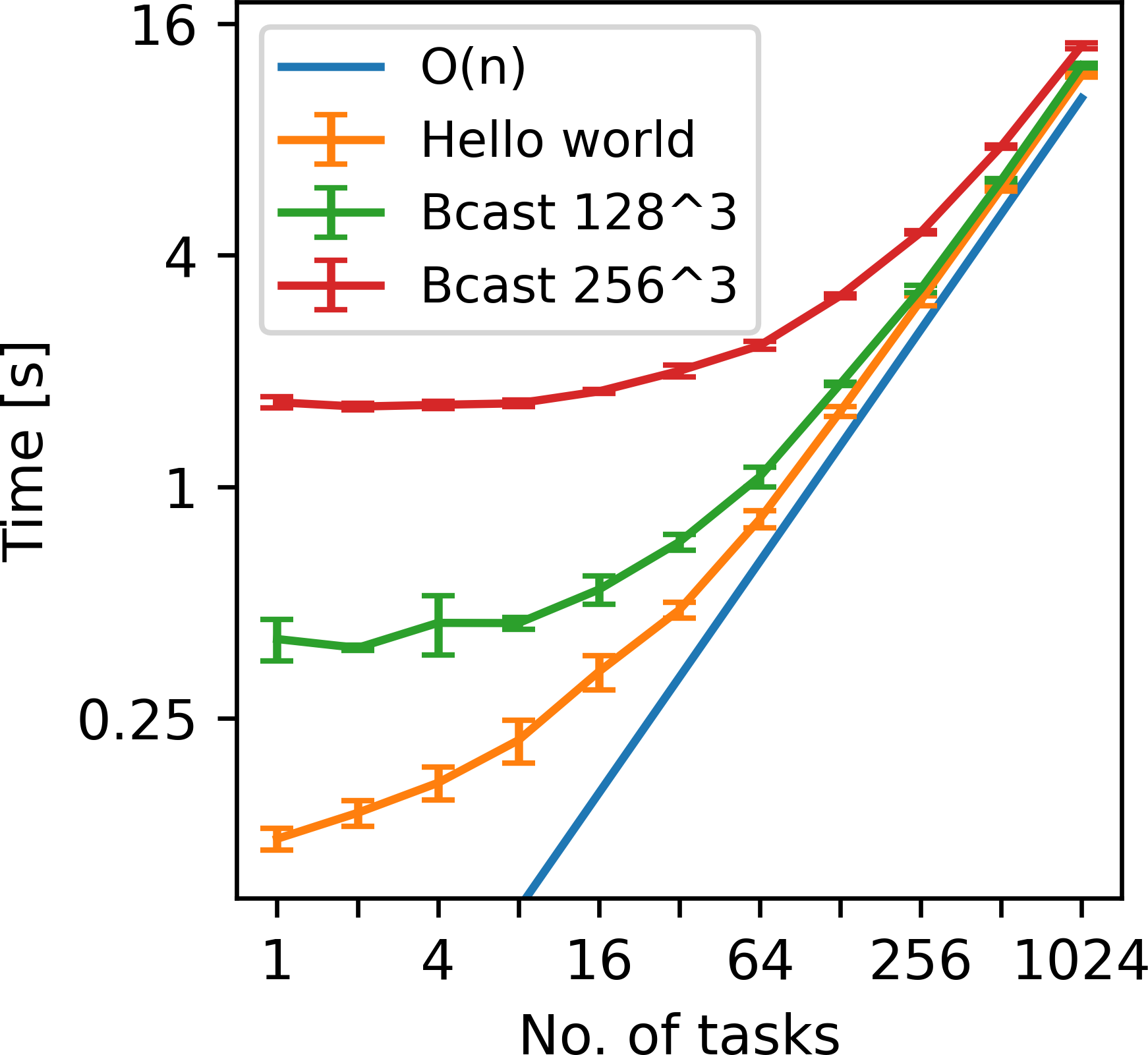}}
\subfloat{\label{fig:redwood-scaling-b}}{\includegraphics[width=0.49\columnwidth]{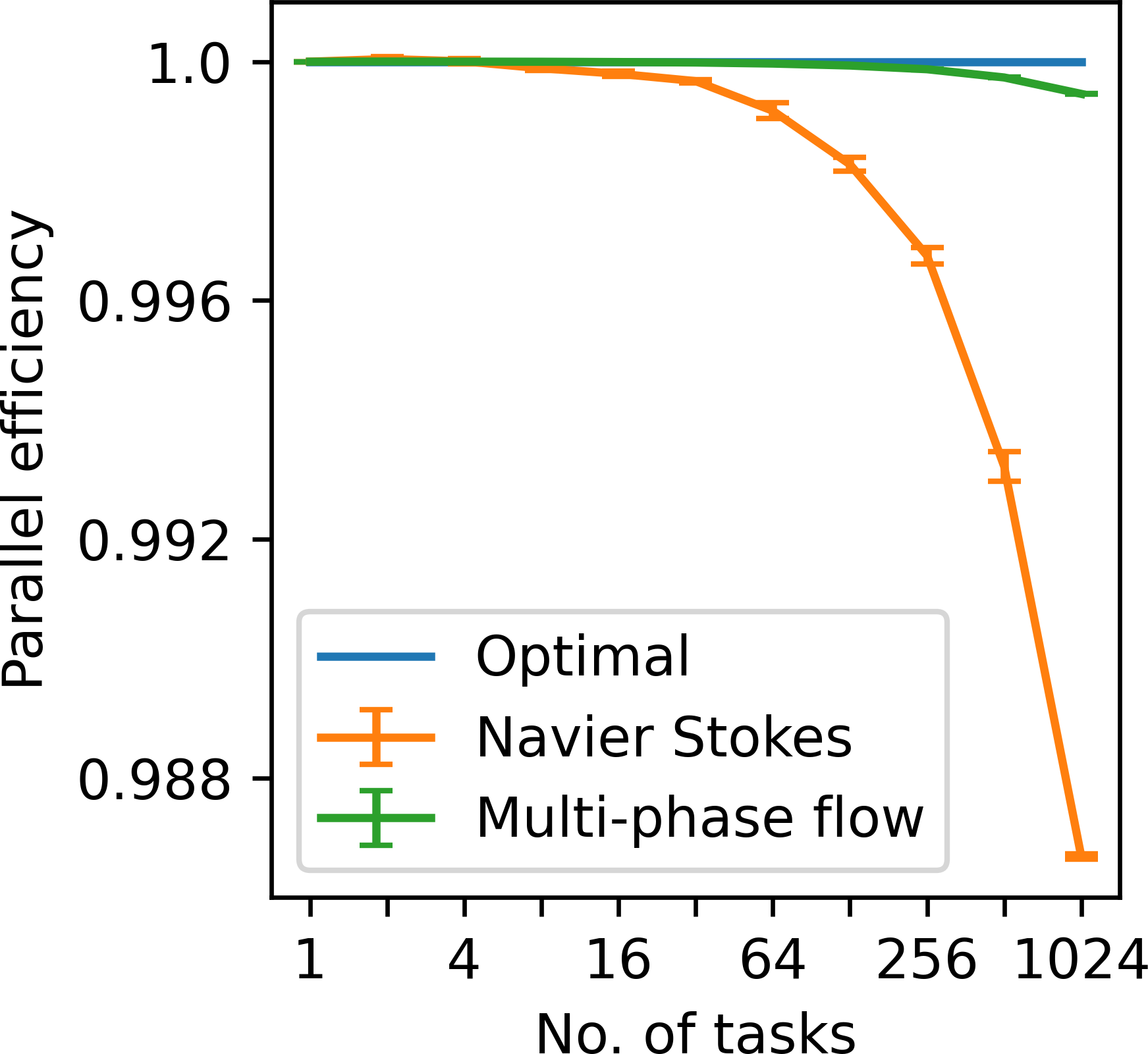}}
\caption{Time to submit an Azure Batch job with Redwood using an increasing number of parallel tasks (left) and weak scaling efficiency for simulating the two datasets from the examples section.}
\label{fig:redwood-scaling}
\end{figure}

We investigate how long it takes to submit a job with an increasing number of tasks by executing a Julia function $n$ times in parallel (using the parallel mapping function).  Submitting tasks to Azure Batch involves Redwood's code generation, as well as the serialization and upload of code and function arguments. As a baseline, we measure the task submission time of an increasing number of invocations of the \textit{hello-world} example from Fig.~\ref{fig:redwood-code}. The results in Fig.~\ref{fig:redwood-scaling-a} show that for a small number of function invocations, task submissions are dominated by the code generation and upload time, which happens only once, regardless of many tasks we submit. However, for more than 16 tasks, the task submission time is dominated by the the time it takes to upload the function argument, which is uploaded $n$ times, as function arguments can be unique to each function invocation. Eventually, the task submission time scales linearly with the number of tasks. 

Next, we test how long it takes to broadcast a 3D Julia array to an increasing number of tasks (running on separate VMs). Redwood's broadcast macro uploads data once to the object store and returns a reference to the data that can be passed as a function argument in place of the original array. Each task then calls the \texttt{fetch} function on the reference to copy the data from blob storage to the worker. The time to submit a job with a small number of tasks is now higher than for the hello-world example, as it includes the time to broadcast the array. However, once we reach a certain number of tasks, the submission time is again dominated by the upload time of the function arguments and thus eventually reaches linear behavior. Broadcasting bigger arrays further increases the job submission time for small number of tasks and shifts the point at which linear behavior sets in to a larger number of tasks.

Our experiments show that, in the worst case, job submission time grows linearly with the number of tasks. One question is whether it is worth to optimize the job submission time, e.g., using recursion. Optimizing the task submission time to reduce latency is important for serverless functions, in which the actual function execution time is small (in the range of milliseconds or seconds). However, with Redwood we specifically target long-running HPC workloads that run on the order of multiple minutes to hours, so we argue that job submission times of e.g., 16 seconds for 1,024 tasks are acceptable. To illustrate this, we compute the parallel weak scaling efficiency for the data generation of our numerical examples. To simulate the training data for our two AI simulators, we solve 3,200 instances of the 3D Navier-Stokes equation and 1,600 instances of the 3D two-phase flow equation. The average task runtime for each scenario are 15 minutes and 6.8 hours respectively. Running these simulations with Redwood is embarrassingly parallel with the only serial component being the task submission. As the task submission time is small in comparison to the overall runtime, both examples reach a high parallel efficiency above 99 percent (Fig.~\ref{fig:redwood-scaling-b}).

\subsection{Neural network architecture}

We choose Fourier Neural Operators (FNOs) as the base architecture for our AI-driven simulator, as they have shown strong performance on a variety of PDEs such as the Navier-Stokes equations or multi-phase flow \cite{li2020fourier, wen2021u}. The parallel FNO implementation based on domain decomposition is introduced in \cite{grady2022towards} and we refer to that paper for implementation details. As we introduce a modification to the original implementation, we include the algorithm of the (updated) parallel FNO implementation in this section. The implementation is based on parallel primitives from the DistDL library \cite{hewett2020linear}.  For FNOs, we rely on the tensor-parallel \textit{broadcast} and \textit{re-partition} primitives. The broadcast primitive is a partition-aware generalization of the classical parallel communication primitive and the re-partition primitive is a generalization of the all-to-all communication pattern for arbitrarily high-dimensional Cartesian data.

First, we establish the mathematical notation. Capital letters represent multi-dimensional tensors and subscripts are dimension labels. We use six-dimensional data tensors $X_{bcxyzt}$ with dimensions batch size $b$, channel $c$, spatial dimensions $x, y, z$ and time $t$. Dimensions in Fourier space are labeled $k_x, k_y, k_z, k_t$. Caligraphic capital letters are linear operations and subscripts indicate the dimensions along which they operate. I.e., $\mathcal{F}_x$ is a Fourier transform along the $x$ dimension and $\mathcal{S}_{yzt}$ represents subsampling or truncation along dimensions $y, z,$ and $t$. Operator $\mathcal{B}$ is the broadcasting operation, which copies a tensor from the master worker to all other workers. $R_{x \rightarrow y}$ is the re-partition operation whose subscripts represent the distributed dimensions before and after partitioning. The partitioned tensor dimension is underlined. For tensor multiplications, we use the Einstein summation notation from PyTorch. Multiplications along dimensions that appear both in the inputs and the output are element-wise multiplications and otherwise multiplications are followed by a summation. E.g., the operation $Y_{bc_oxyzt} = X_{bc_ixyzt} W_{c_ic_oxyzt}$ performs an element-wise multiplication along dimensions $xyzt$ (which appear in all three tensors) and a multiplication followed by a sum along the input channel dimension $c_i$ (which only appears on the right-hand side). Letters $W_{c_ic_oxyzt}$ are network weights.

\begin{algorithm}[bt]
\caption{Architecture of Model-Parallel FNO.}\label{alg:fno}
\texttt{\# Encoder} \\
$W^e_{c_ic_o} \leftarrow \mathcal{B} W^e_{c_ic_o}$ \\
$X_{bc\underline{x}yzt} \leftarrow \sigma(X_{bc\underline{x}yzt} W^e_{c_ic_o})$ \\

\texttt{\# FNO blocks (i=1,2,3,4)} \\
$X_{bc\underline{x}yzt}^{i+1} \leftarrow \sigma \Big( fno\_block \big(X^i_{bc\underline{x}yzt}, W^i_{c_ic_ok_x\underline{k_y}k_zk_t}\big) \Big)$\\

\texttt{\# Decoder} \\
$W^d_{c_ic_o} \leftarrow \mathcal{B} W^d_{c_ic_o}$ \\
$X_{bc\underline{x}yzt} \leftarrow \sigma(X_{bc\underline{x}yzt} W^d_{c_ic_o})$
\end{algorithm}

The model-parallel FNO implementation based on this notation is shown in Algorithm~\ref{alg:fno} and closely follows the architecture of the original FNO  \cite{li2020fourier}. The network consists of an encoder that increases the channel dimension of the input through a one-by-one convolution. This is followed by a number of FNO blocks, which are separately defined in Algorithm~\ref{alg:fno-block} and a decoder that brings the channel dimension down to the desired number of output channels. In the model-parallel FNO version, the input tensor $X$ is distributed across the first spatial dimension $x$. As the convolutions in the encoder and decoder do not sum along this dimension, we simply need to broadcast the encoder/decoder weights during the forward pass and perform the tensor multiplications independently on each worker. 

\begin{algorithm}
\caption{Architecture of Distributed FNO Block.}\label{alg:fno-block}
\texttt{\# Distributed FFT and freq. truncation} \\
$X_{bck_x\underline{k_y}k_zk_t} \leftarrow \mathcal{S}_{x} \mathcal{F}_{x} \mathcal{R}_{x \rightarrow y}\mathcal{S}_{yzt} \mathcal{F}_{yzt} X_{bc\underline{x}yzt}$ \\

\texttt{\# Spectral convolution} \\
$X_{bc_ok_x\underline{k_y}k_zk_t} \leftarrow X_{bc_ik_x\underline{k_y}k_zk_t} W_{c_ic_ok_x\underline{k_y}k_zk_t}$ \\

\texttt{\# Padding and inverse FFT} \\
$X_{bc\underline{x}yzt} \leftarrow \mathcal{F}_{yzt}^\top \mathcal{S}_{yzt}^\top  \mathcal{R}_{x \rightarrow y}^\top \mathcal{F}_{x}^\top \mathcal{S}_{x}^\top X_{bc_ok_x\underline{k_y}k_zk_t}$
\end{algorithm}

\begin{figure}[b]
\centerline{\includegraphics[width=0.99\columnwidth]{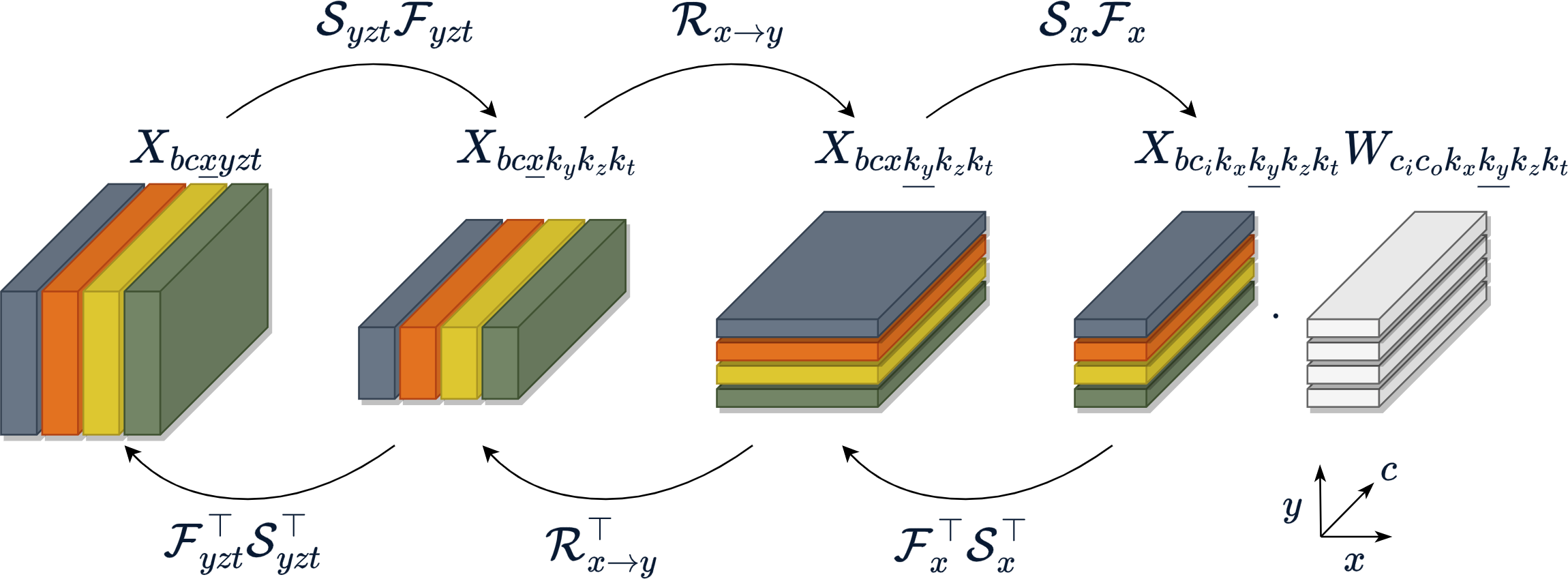}}
\caption{Distributed spectral convolution of the FNO block. We first compute the FFT along the non-partitioned dimensions and truncate the high frequency. Then, data is re-partitioned and we compute the FFT along the final dimension. After multiplication with the learnable weights, we repeat the operations in reverse order using the adjoint of the operators.}
\label{fig:fnoblock}
\end{figure}

The FNO blocks perform spectral convolutions in the Fourier domain and compute 4D Fourier transforms (FFT) of the input along the spatial-temporal tensor dimensions ($xyzt$). As in the original FNO, most frequencies are truncated after the FFT to reduce the number of learnable weights. In the model-parallel version (Algorithm~\ref{alg:fno-block}), the input tensor is initially partitioned along the spatial $x$ dimension, so we cannot directly compute the 4D FFTs. We first compute a 3D FFT along the non-partitioned dimensions and apply frequency truncation along those dimensions to reduce the data size (Fig.~\ref{fig:fnoblock}). Next, we apply the re-partition operator, which distributes data along the $y$ dimension and we can compute the final FFT along the $x$ dimension. The tensor multiplication with the weight tensor is an element-wise multiplication in the spatial-temporal dimensions ($k_xk_yk_zk_t$) and summation only occurs along the (non-partitioned) channel dimension. Each worker therefore maintains its own portion of weights and no communication is required for the spectral convolution itself. For the inverse FFT, we apply the same steps as before in reverse order, using the adjoint (conjugate transpose) of the linear operators.

The parallel FNO version in \cite{grady2022towards} uses a two-dimensional partitioning scheme and performs frequency truncation after the re-partitioning. This results in a total of four re-partition operations per FNO block, during each of which we communicate the full tensor $X$. Algorithm \ref{alg:fno-block} performs only two re-partition operations per FNO block, during each of which we only communicate a tensor whose size has been truncated along three dimensions. In our examples, we truncated around 80 percent of the frequencies in each dimension, thereby reducing the amount of communicated data by a factor of 160 per re-partition operation.

\subsection{Network performance}

The performance evaluation of our parallel FNO implementation is motivated by our goal to scale AI simulators to industry-scale problem sizes. To advance the current state of the art, we do not need to scale model-parallelism across hundreds of GPUs on a large network, but rather reach high parallel efficiency on a small number of GPUs on a single node with high-bandwidth GPU interconnect. The current version of DistDL uses an MPI communication backend, which includes support for cuda-aware MPI. However, to achieve the best possible performance, we implement a NCCL backend for DistDL primitives. As NCCL is optimized for Nvidia's GPU topologies, this ensures that we can take optimal advantage of Nvidia's NVLink interconnect. All scaling tests are performed on a single Azure ND96amsr VM with eight Nvidia A100 GPUs, each with 80 GB of RAM.

We select a problem setup that occupies about 80\% of the memory on a single GPU and we use randomly generated input data of batch size one. We are interested in increasing the spatial dimension of the overall problem by using additional GPUs, while keeping the problem size \textit{per} GPU fixed (i.e., weak scaling). We increase both the size of the input data and the number of weights, so both the memory footprint, as well as number of floating point operations (FLOPs) per GPU stays constant. As we are using data with a batch size of one, we cannot use data parallelism and ZeRO, both of which require at least a batch size equal to the number of GPUs. This leaves us with pipeline parallelism, which allows us to partition the network and data across multiple GPUs, although it does not provide any concurrency for a batch size of one.

\begin{figure}[bt]
\subfloat{\label{fig:weakscaling-a}}{\includegraphics[width=0.48\columnwidth]{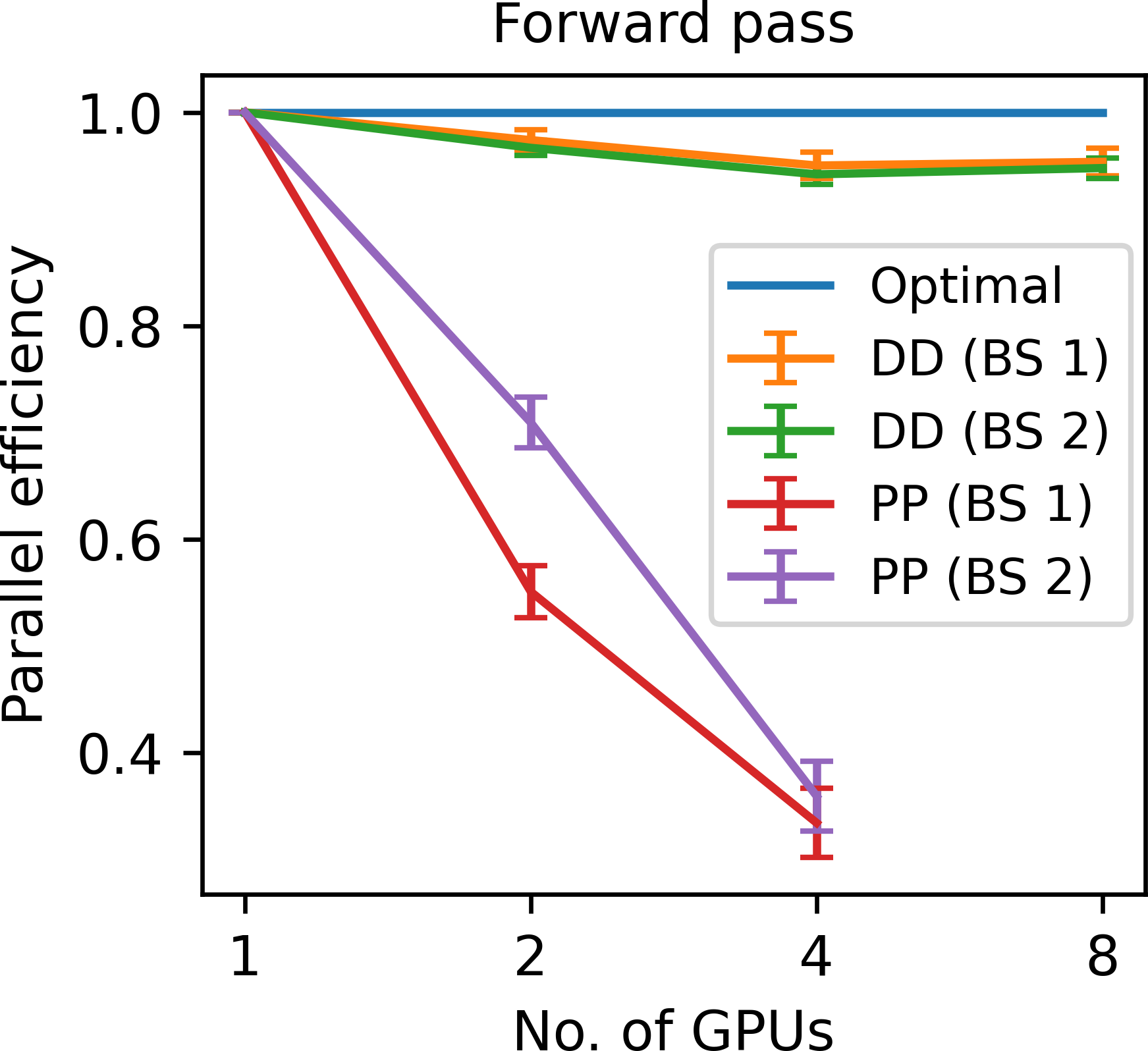}}
\hspace*{.1cm}
\subfloat{\label{fig:weakscaling-b}}{\includegraphics[width=0.48\columnwidth]{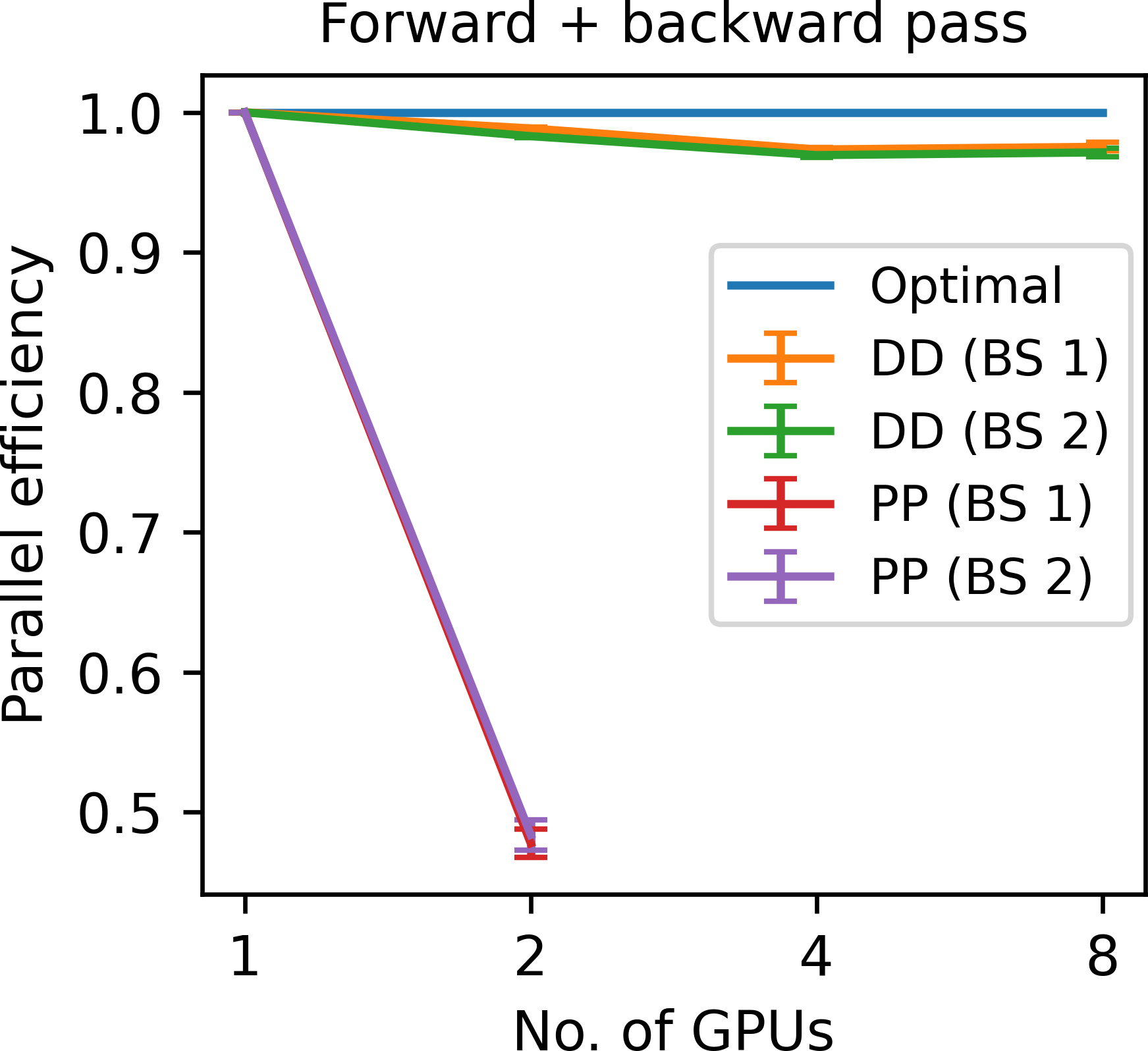}}
\caption{Weak scaling of model-parallel FNOs with pipeline parallelism (PP) and domain decomposition (DD), using data with batch sizes (BS) of one and two. Both the memory footprint and number of FLOPs per GPU are constant. Error bars indicate the 95\% confidence interval over 16 runs.}
\label{fig:weakscaling}
\end{figure}

Our weak scaling experiments (Fig.~\ref{fig:weakscaling}) confirm this, as we reach 50\% parallel efficiency on 2 GPUs for our pipeline parallel FNO (PP) and 25\% on 4 GPUs (i.e., no concurrency). In contrast, the FNO based on domain decomposition (DD) achieves above 90\% parallel efficiency in the forward pass and above 95\% in forward- plus backward pass. On eight GPUs (the maximum number of GPUs in a single virtual machine), pipeline parallelism runs out of memory, even though the problem size per GPU is fixed, which indicates that PyTorch's pipeline parallelism module suffers from a memory overhead. When computing the backward pass as well, pipeline parallelism runs out of memory for more than 2 GPUs. As pipeline parallelism relies on larger batch sizes to achieve concurrency, we repeat the scaling experiments for a batch size of two as well (by making the spatial dimension smaller so that the memory footprint stays the same). On 2 GPUs, pipeline parallelism now achieves in fact some level of concurrency (parallel efficiency larger than 50\%), but on 4 GPUs the efficiency decreases again (likely because the batch size is smaller than the number of GPUs). Domain decomposition reaches high parallel efficiency in both cases, thereby demonstrating the strengths of this approach, which does not rely on a specific batch size to reach high levels of concurrency.

\begin{figure}[tb]
\subfloat{\label{fig:strongscaling-a}}{\includegraphics[width=0.48\columnwidth]{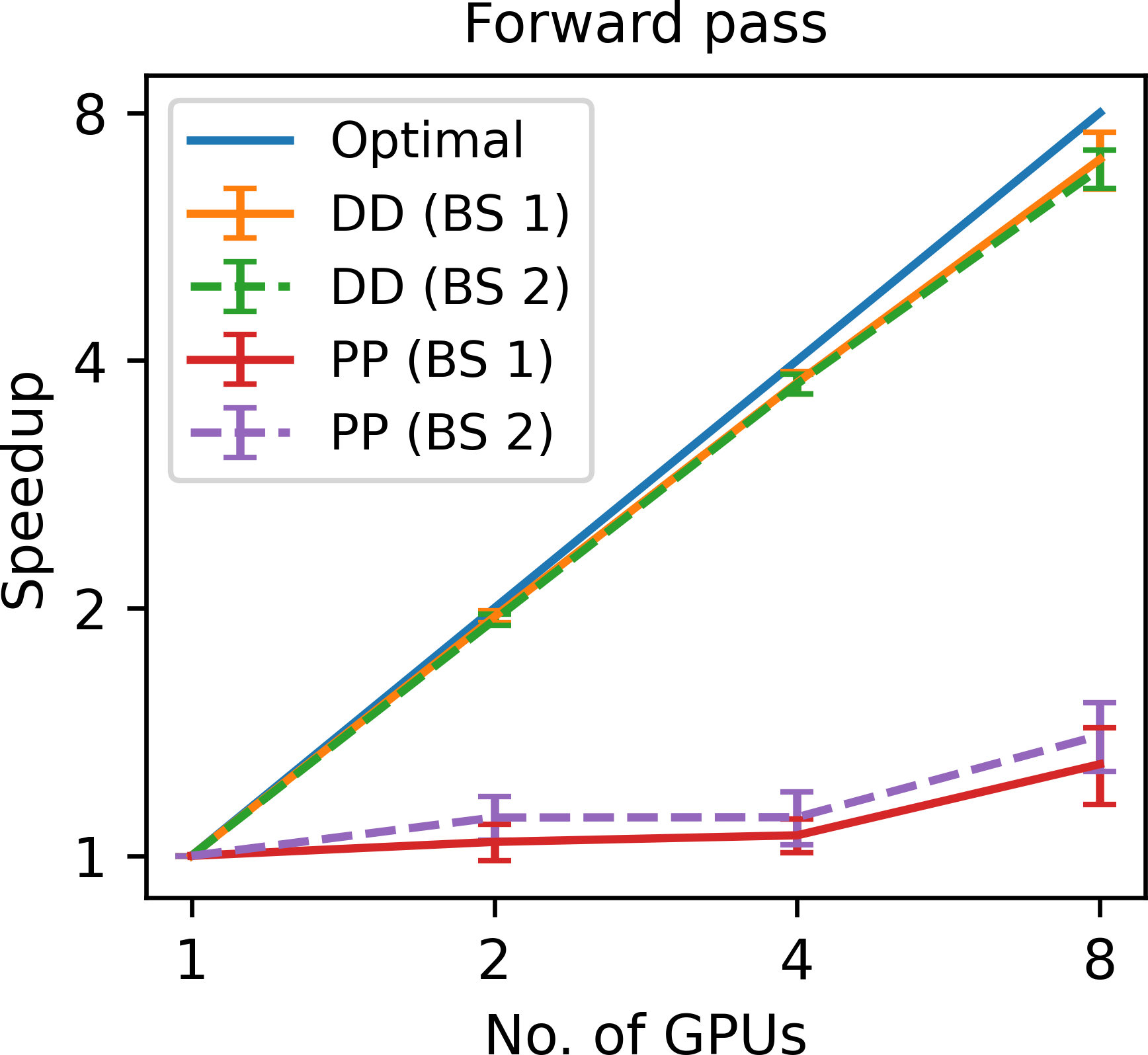}}
\hspace*{.1cm}
\subfloat{\label{fig:strongscaling-b}}{\includegraphics[width=0.48\columnwidth]{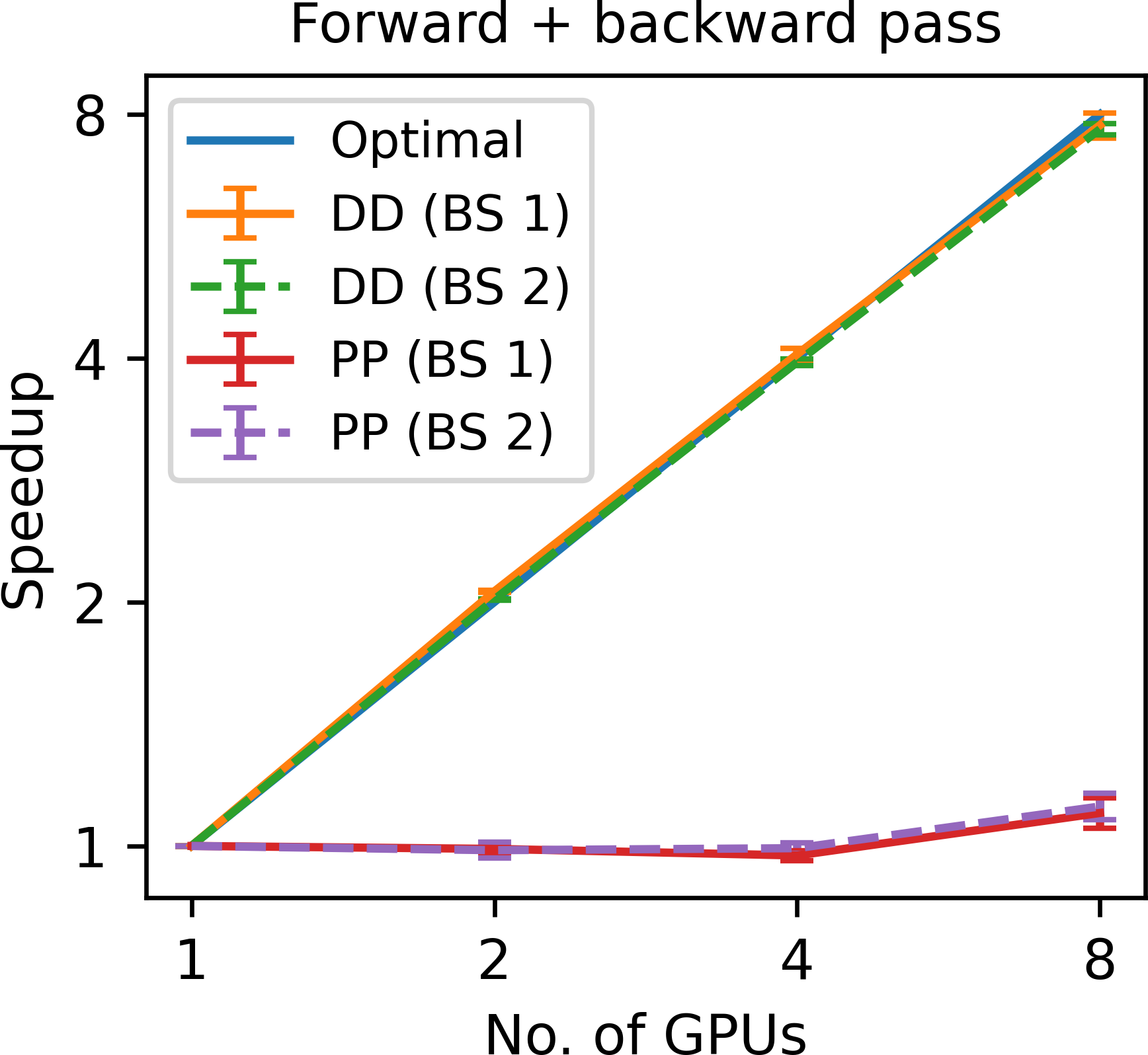}}
\caption{Strong scaling of model-parallel FNOs with pipeline parallelism (PP) and domain decomposition (DD) using varying batch sizes (BS). The memory footprint and number of FLOPs on each GPU is reduced according to the total number of GPUs. Error bars represent the 95\% percent confidence interval over 16 runs.}
\label{fig:strongscaling}
\end{figure}

While strong scaling is less relevant for training neural networks, as memory usage usually dictates the number of GPUs required, it is important for speeding up inference time. We therefore investigate the strong scaling behavior of pipeline parallelism and domain decomposition as well. We use the same problem setup as in the previous experiment, but keep the overall data dimensions fixed so that the problem size per GPU shrinks. Once again, we reach high performance and nearly linear scaling with domain decomposition and generally poor performance with pipeline parallelism (Fig.~\ref{fig:strongscaling}).

\section{Applications}

Using the two tools presented in this paper, we train two AI-based numerical simulators for solving large-scale 3D PDEs with deep learning. We advance the current state of the art in scientific AI by a factor of eight in terms of the number of predicted variables, which correspond to the grid or mesh size and number of predicted time steps. For both examples we train 4D FNOs that predict spatial-temporal solutions of PDEs with more than 140 million output variables per sample. Once trained, these AI surrogate models are valuable in downstream applications such as optimization or uncertainty quantification (which are outside the scope of this paper).

\subsection{Turbulent flow}

In our first example, we train an FNO-based surrogate model for simulating turbulent flow around a sphere by solving the 3D Navier-Stokes equation. In our dataset, we vary the location of the sphere in three-dimensional space, which leads to different flow patterns, depending on the location and distance of the sphere from the model edges (using Dirichlet boundary conditions). We generate 3,200 data pairs consisting of input and output data. Each input is a 3D binary map that indicates the location of the sphere and each output is a 4D tensor of the simulated vorticity of dimensions 130 $\times$ 130 $\times$ 130 $\times$ 64 (three spatial dimensions plus time). As FNOs require that inputs and outputs have the same dimensions, the input binary map is repeated along the time dimension.

\begin{figure}[bt]
\subfloat{\label{fig:ns-datagen-a}}{\includegraphics[width=0.48\columnwidth]{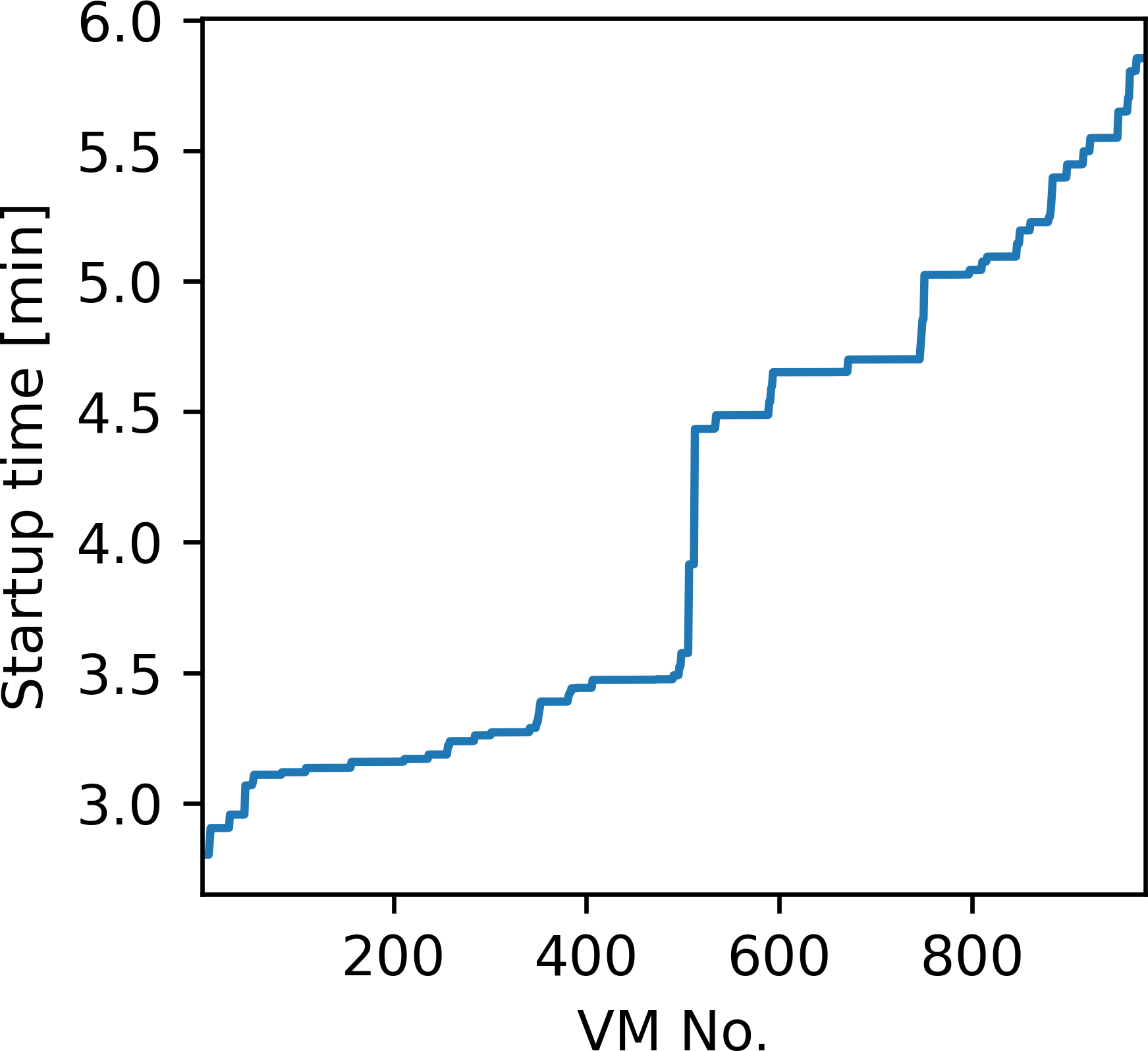}}
\hspace*{.1cm}
\subfloat{\label{fig:ns-datagen-b}}{\includegraphics[width=0.48\columnwidth]{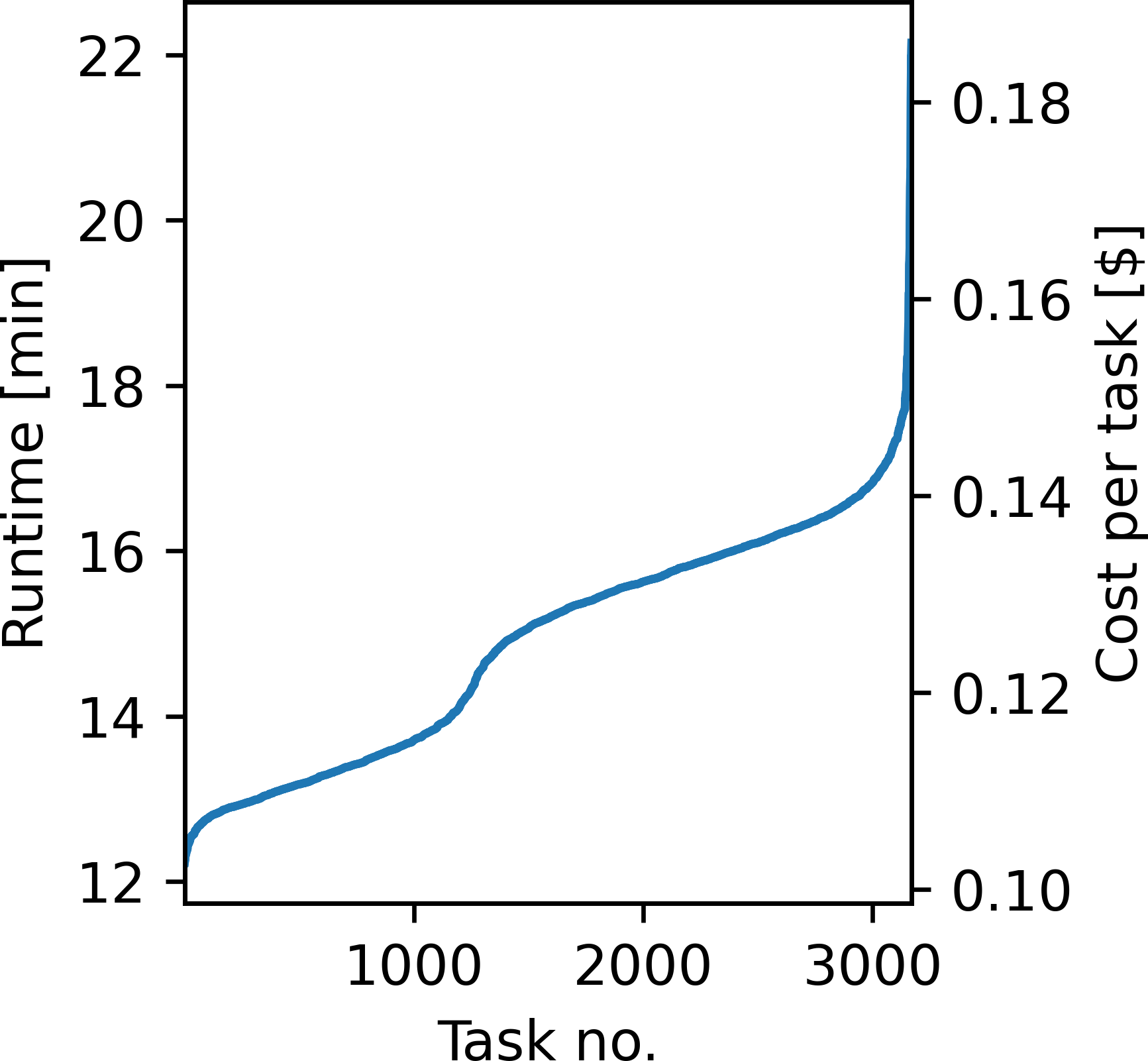}}
\caption{Startup time of VMs from the batch pool and (sorted) runtimes of the 3,200 Julia tasks for simulating the training data. The average task runtime is 15 minutes.}
\label{fig:ns-datagen}
\end{figure}

We simulate the training data with \textit{WaterLily.jl}, an open-source Julia package for solving the 2D and 3D Navier-Stokes equations with the geometric multigrid method \cite{weymouth2021}. We implement a Julia function that takes the location of the sphere as input, solves the 3D Navier-Stokes equation with WaterLily, and outputs the scalar vorticity as a function of space and time (i.e. as a 4D tensor). Using Redwood, we create a batch pool of 1,000 Azure VMs (E4s v3 with 4 vCPU cores) and run 3,200 simulations to generate the training data. Fig.~\ref{fig:ns-datagen-a} shows the time it takes to launch the 1,000 VMs. About half of the VMs are available after 3.5 minutes and most remaining VMs are available after 6 minutes. Note that Azure Batch starts scheduling tasks as soon as the first VMs become available, so users do not have to wait for all VMs to spin up first. Fig.~\ref{fig:ns-datagen-b} shows the runtime of each of the 3,200 tasks and the cost of simulating each training sample. The average simulation time is 15 minutes per sample and the cost for generating the full dataset on Azure is \$396 with on-demand VMs and \$158 using spot VMs \cite{pricing2022}. Each task writes its simulated training pair to Azure Blob storage using Zarr \cite{miles2020}, a Python package for storing N-dimensional arrays on various storage backends, including cloud object storage. Each training sample has a size of 536 MB and the total data set is 1.6 TB (in uncompressed form). 

We train the FNO on a single Azure ND96amsr VM, the same VM as used in the performance evaluation. We train for 50 epochs using a batch size of two, which is the maximum possible batch size before running out of memory. We use 2,800 of the data samples for training and validation and save 400 samples for testing. The training time per epoch is around 30 minutes and total training time is close to 24 hours (on-demand price of \$786 and spot price of \$393) \cite{pricing2022}. As the input for the FNO is partitioned along the first spatial dimension, each GPU reads its corresponding chunk of the data from blob storage during the first training epoch. The full dataset (1.6 TB) approaches the limits of the VM's CPU memory (1.9 TB), so we cache the training data on a local NVMe drive from which we re-read the data during subsequent training epochs.

\begin{table}[b]
  \caption{Performance on Validation and Test Data (MSE: Mean Squared Error, MAE: Mean Average Error).}
  \label{table:scores}
  \centering
  \begin{tabular}{lccc}
    \toprule
         & MSE  &  MAE & R2 \\
    \midrule
    Navier-Stokes: Validation & $0.0552$ & $0.5851$ & $0.9714$  \\
    Navier-Stokes: Test     & $0.0507$ & $0.5587$ & $0.9734$ \\
    \midrule
    \CO flow: Validation & $1.1104 \cdot 10^{-4}$ & $0.0866$ & $0.9453$\\
    \CO flow: Test     & $1.1603 \cdot 10^{-4}$ & $0.0952$ & $0.9487$ \\
    \bottomrule
  \end{tabular}
\end{table}

\begin{figure}[bt]
\centerline{\includegraphics[width=0.99\columnwidth]{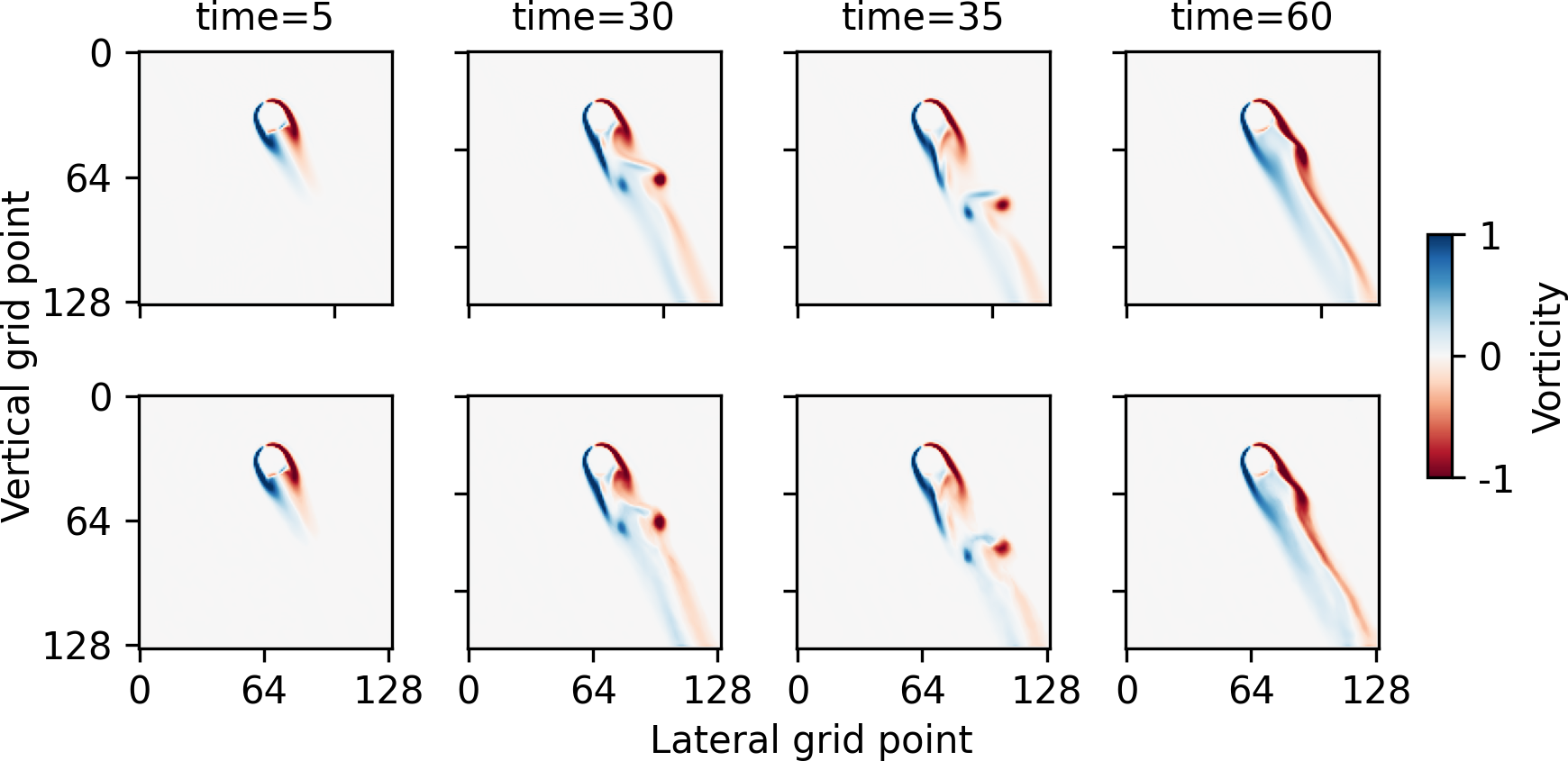}}
\caption{2D slices through the 3D vorticity at different time steps as simulated with WaterLily and as predicted by the model-parallel FNO. The simulation time with WaterLily is around 15 minutes (on 4 vCPUs) and 10 ms with the FNO (on 8 A100s).}
\label{fig:ns-results}
\end{figure}

Network performance on the validation and test data is listed in Table~\ref{table:scores}. Fig.~\ref{fig:ns-results} shows several 2D slices of the predicted data at different time steps in comparison to the data simulated with WaterLily. The sphere location shown in the example was drawn from the test dataset and was not seen by the network during training. The FNO is able to predict the vorticity in .1 seconds on 8 A100s, whereas the numerical simulation with WaterLily takes around 15 minutes on 4 CPU cores. Taking into account the cost differences of the VMs, we arrive at a cost of 6.25 cents per simulation with WaterLily on the E4s VM and 0.09 cents per simulation using the FNO on the ND96amsr VM. If we account for the cost of data generation and training, the FNO amortizes that cost after 19,188 simulations and will save money for any additional simulations. As downstream applications such as optimization potentially require tens of thousands of (sequential) simulations, this opens up both cost and time savings.

\subsection{CO$_2$ flow}

In our second example, we train an FNO for simulating \CO{} flow in an industry-scale carbon capture scenario. For simulating the training data, we use the Sleipner 2019 benchmark model \cite{sleipner2019}, a real-world geological model for 3D numerical reservoir simulations. Sleipner is the world's first commercial CCS project and located off the coast of Norway in the North Sea. The benchmark model simulates the \CO{} plume behavior as observed during the project, which used a single \CO{} injection well \cite{singh2010reservoir}. To train our FNO, we simulate training data with the original Sleipner geomodel, but using multiple concurrent \CO{} injection wells that vary spatially. At test time, we predict \CO{} flow at new well locations for up to four wells. Note that we do not subsample the original model and use the full simulation grid of size 262 $\times$ 118 $\times$ 64 for training and testing. Our FNO is trained to predict the \CO{} saturation history for 86 time steps, which results in a total of 170 million output variables.

\begin{figure}[bt]
\subfloat{\label{fig:co2-datagen-a}}{\includegraphics[width=0.48\columnwidth]{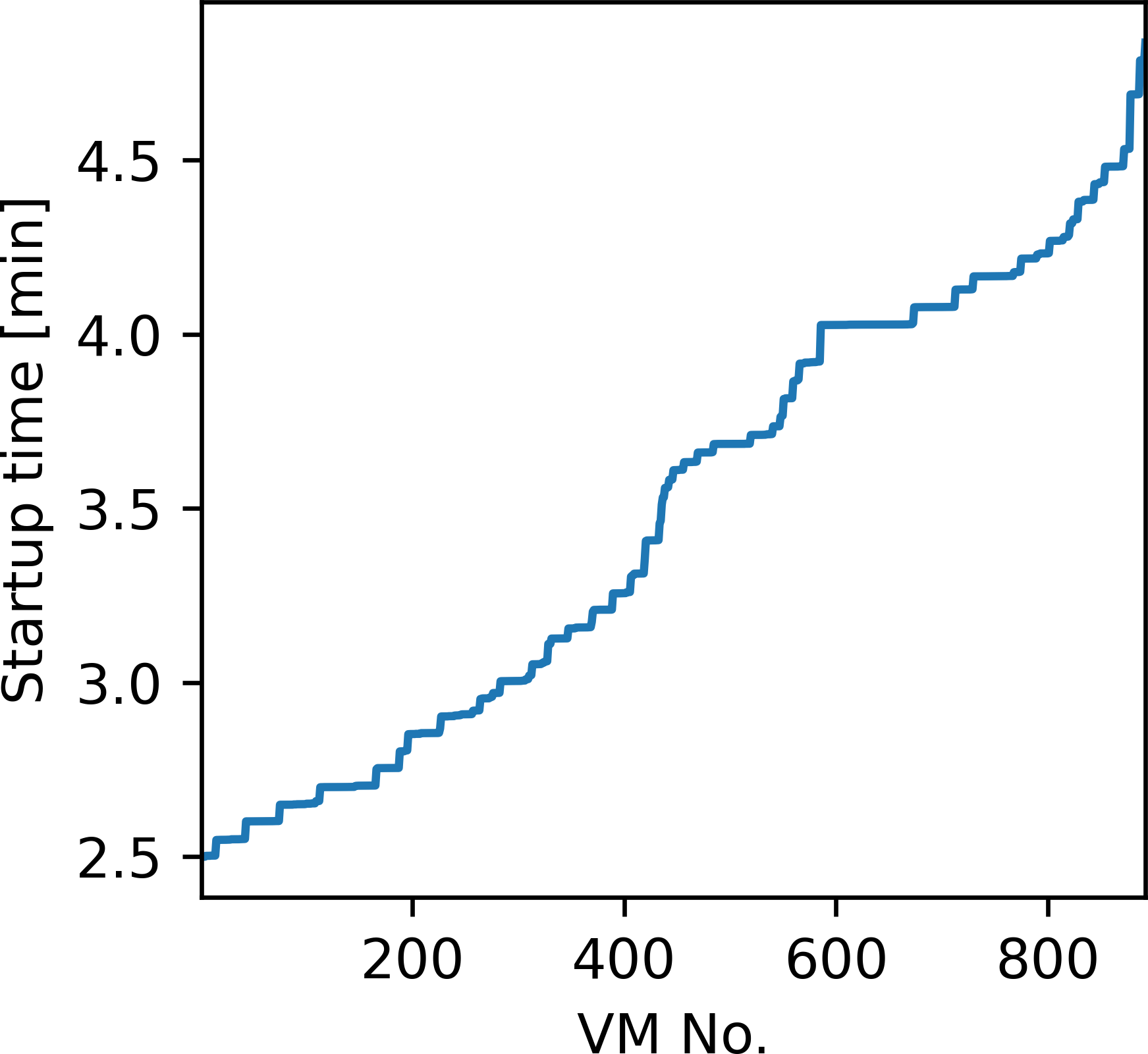}}
\hspace*{.1cm}
\subfloat{\label{fig:co2-datagen-b}}{\includegraphics[width=0.48\columnwidth]{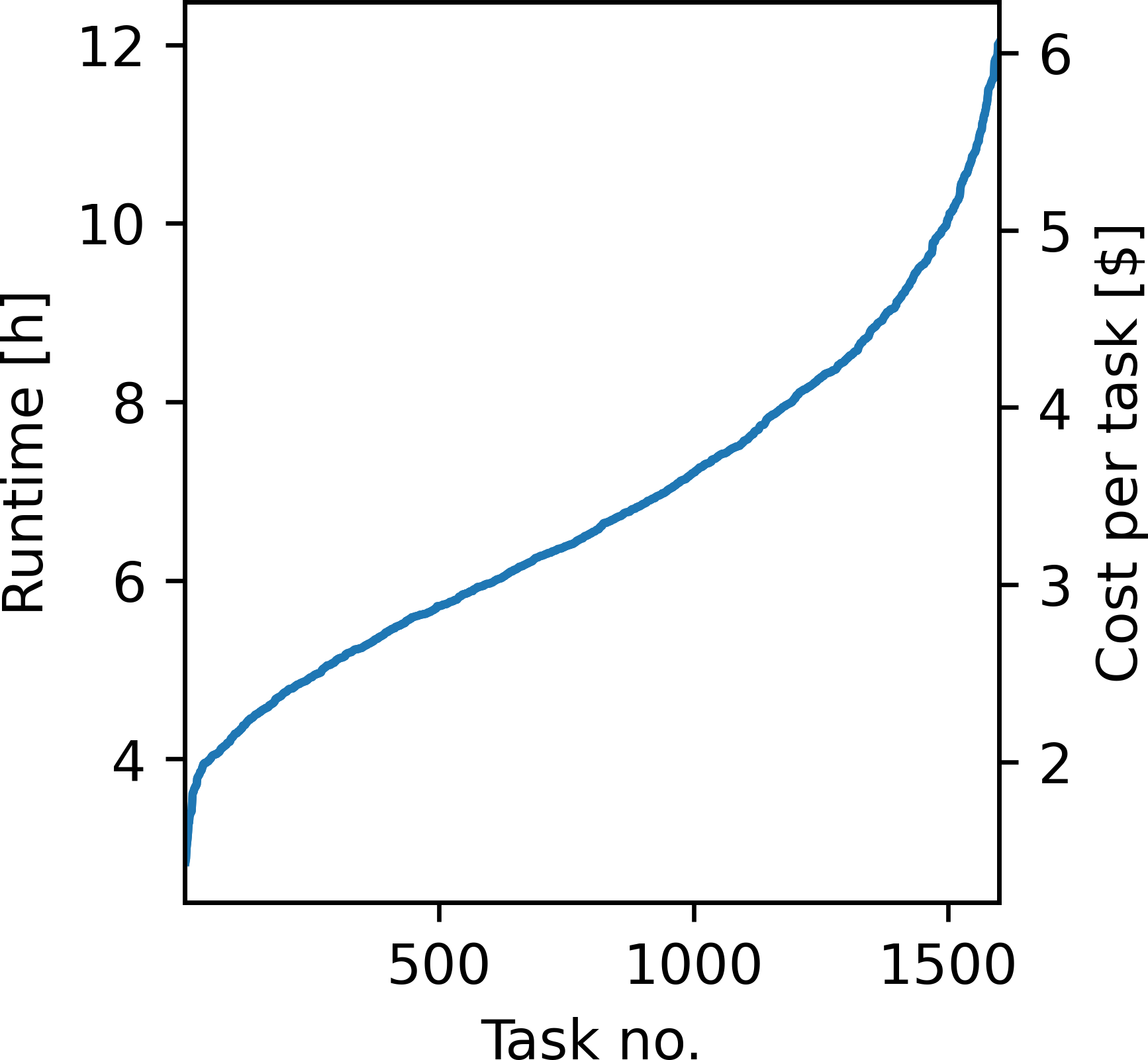}}
\caption{VM startup times for the Sleipner data simulation and (sorted) task runtimes. The average simulation time is 6.8 hours.}
\label{fig:co2-datagen}
\end{figure}

For training, we simulate 1,600 data samples with the Open Porous Media (OPM) simulator \cite{rasmussen2021open}, an open-source reservoir simulator written in C++ and based on the finite volume method. We use the simulator configuration and parameters from the Sleipner benchmark and only change the number and location of injector wells. Even though OPM is not written in Julia, we can still use Redwood for the training data generation. We set up a docker image with OPM and the Redwood runtime, which is automatically deployed to the VMs by Azure Batch. Using Redwood, we write a Julia function that runs the simulator on each worker, reads the simulated output back into Julia, and stores it in blob storage for training. As before, we use a batch pool with 1,000 Azure VMs (E8s with 8 vCPUs). The VM startup times and runtime of each task are shown in Fig.~\ref{fig:co2-datagen}. Compared to the Navier-Stokes example, the simulation time per sample is much larger (6.8 hours on average) and the total cost for data generation is \$5,487 with on-demand VMs and \$2,194 with spot VMs \cite{pricing2022}.

\begin{figure}[bt]
\centerline{\includegraphics[width=0.99\columnwidth]{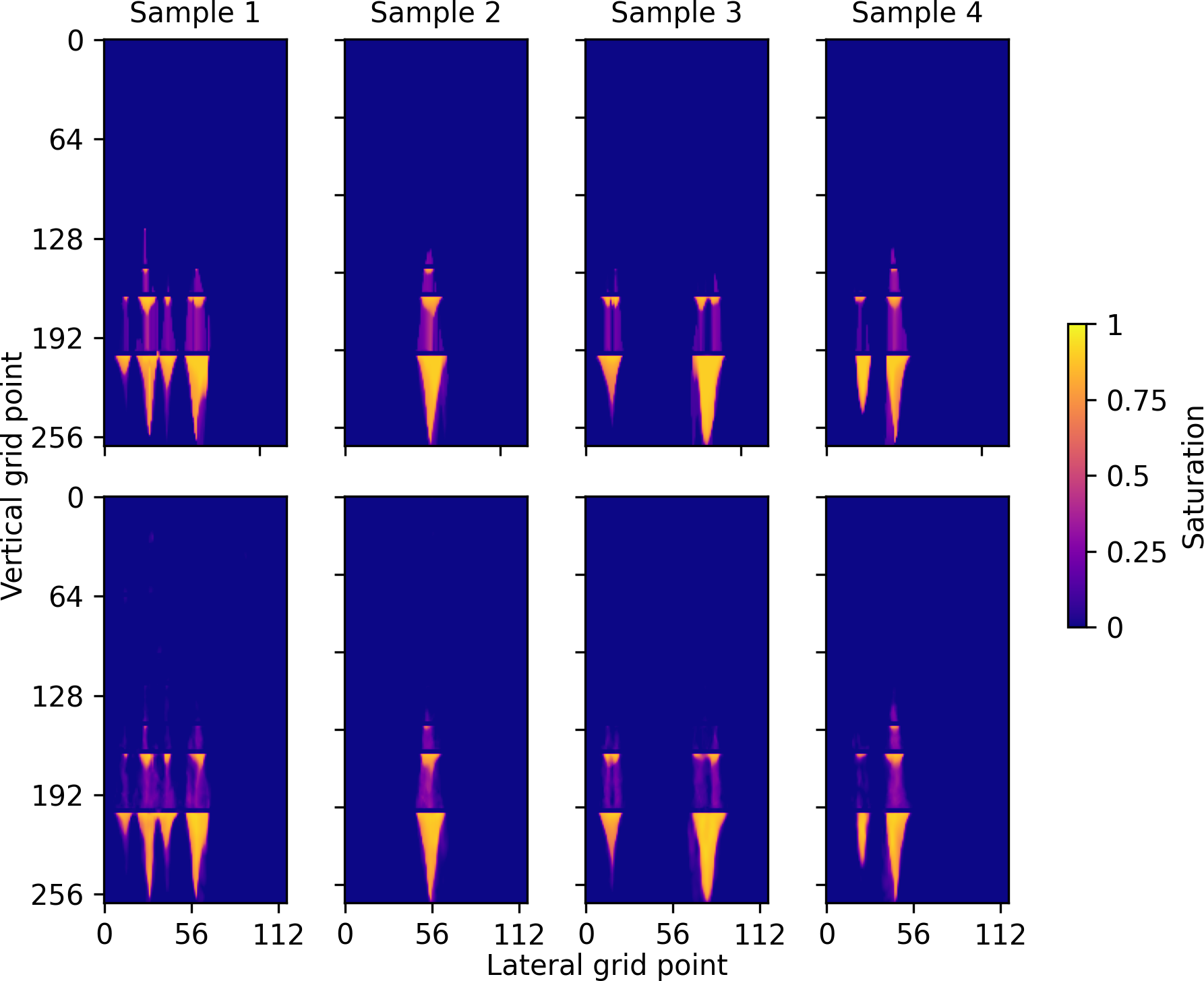}}
\caption{\CO{} saturation at the final simulation time step as predicted by OPM (top) and the FNO (bottom). Each column shows a 2D slice through the 4D data tensor for different locations and numbers of \CO{} injection wells.}
\label{fig:co2-results}
\end{figure}

Each of the 1,600 training pairs consists of a 3D binary map that indicates the locations of the injection wells (repeated along the time dimensions) and the simulated \CO{} saturation history as a 4D tensor of space and time. As before, we train on a single Azure node with 8 A100s for 50 epochs and a batch size of two. The training time per epoch is around 20 minutes and the total training time is 17 hours (\$557 on-demand and \$279 spot price) \cite{pricing2022}. Fig.~\ref{fig:co2-results} shows several 2D slices (from the 4D volumes) at the final time step for four different well scenarios as modeled with OPM (top row) and the FNO (bottom row). Simulations with the trained FNO take around .12 seconds (on the ND96amsr VM), whereas the average simulation time with OPM on the E8s VM is 6.8 hours. Adjusting for the difference in VM prices, this results in \$3.4 per simulation with OPM and 0.11 cents per simulation with the FNO (a factor of 3,200). Considering the cost of training data generation and training itself, the FNO breaks even after running 1,848 simulations and is 3,200 times cheaper for any additional simulations. Considering the large number of possibilities to place up to four wells in the model (over 12 billion combinations), the FNO provides a fast and low-cost surrogate model for optimizing well placement or uncertainty quantification.

\section{Discussion}

Model parallelism is more tightly coupled than data parallelism because we communicate data at each neural network layer and not just to synchronize gradients at the end of a forward-backward pass. The high parallel efficiency we obtain on up to 8 GPUs relies on the high-bandwidth NVLink interconnect between GPUs on a single node (up to 600 GB/s). The combined memory of 8 A100s is sufficient to predict 84 times steps on a simulation grid with 2 million cells, which is a total of 170 million output variables. If we use the same configuration to solve a stationary PDE (or predict only one time step at a time), we can process $512^3$ grid points in 3D or $12,000^2$ points in 2D. While these problem sizes are sufficient for many commercial settings, there are cases that require even larger meshes \cite{dogru2011new}, which in turn require scaling model-parallel networks \textit{across} nodes and likely leads to lower parallel efficiency. A limitation of the current implementation is that we solely use model/domain parallelism and train with a small batch size of two. Training with larger batch sizes requires hybrid parallelism models, similar to the strategy from the Turing NLG model, a transformer that combines ZeRO data parallelism with tensor decomposition and pipeline parallelism \cite{smith2022using}.

Both Redwood and DistDL (for implementing the model-parallel FNO) raise the abstraction level for users, who do not have to interact with cloud vendor-specific REST APIs or message passing APIs. Nevertheless, implementing model parallelism with DistDL involves considerably more code changes than implementing data or ZeRO parallelism in PyTorch (tens of lines versus a few lines). To implement model/tensor parallelism for the FNO, it is necessary to partition each tensor of the network manually, including weights/biases, input/output tensors and hidden states. There are many possibilities for choosing a tensor partitioning scheme and the choice of partitioning significantly influences the amount of data communication and performance. E.g., in the FNO implementation, we distribute tensors along one of the spatial-temporal dimensions, which only requires the communication of hidden states during the 4D FFT and iFFT, but not during the encoder and decoder. However, partitioning data tensors along the channel dimension is also possible, in which case the FFTs become embarrassingly parallel, but the encoder and decoder require communication.

\section{Conclusion}

Scientific AI has the potential to address challenging scientific computing problems that rely on expensive numerical simulations, but scaling to commercial problem sizes is the main road block in adopting this technology for industry applications. We introduce an API for simulating large-scale training datasets in the cloud, which in combination with model parallelism for deep learning, enables us to train commercial-scale simulators for solving the Navier-Stokes and two-phase flow equations. We show that using only commodity cloud VMs and a small number of GPUs, we can train the largest AI simulator to date on a real-world reservoir simulation benchmark and unlock scientific AI for commercial-scale settings.

\section*{Acknowledgments}

We thank Thomas Grady from the Georgia Institute of Technology for his contributions to DistDL and the initial work on distributed FNOs with his Ph.D. advisor Felix J. Herrmann. Many thanks also to Erik Skjetne and his colleagues from Northern Lights for sharing their expertise on CCS. We also thank John Godlewski from SLB for all the insightful discussions around reservoir simulations.

\bibliographystyle{IEEEtran}
\bibliography{references}

\end{document}